\documentclass[reprint,superscriptaddress,amsmath,amssymb,aps,prb]{revtex4-2}

\usepackage{graphicx}
\usepackage{dcolumn}
\usepackage{bm}
\usepackage{physics}
\usepackage{amsmath}
\usepackage[english]{babel}
\makeatletter
\providecommand{\l@en}{\l@english}
\makeatother
\usepackage{hyperref}
\usepackage{cleveref}
\usepackage{braket}
\usepackage{float}
\crefname{equation}{Eq.}{Eqs.}
\usepackage[mathlines]{lineno}

\usepackage{xcolor}
\usepackage{letltxmacro}
\definecolor{citebg}{RGB}{255,255,255}
\definecolor{citefg}{RGB}{221,38,127} 

\LetLtxMacro{\oldcite}{\cite}
\renewcommand{\cite}[1]{\colorbox{citebg}{\textcolor{citefg}{\oldcite{#1}}}}
\hypersetup{
    pdfborder={0 0 0}
}

\begin{document}

\definecolor{tolpurple}{RGB}{112,0,254}
\LetLtxMacro{\oldref}{\ref}
\renewcommand{\ref}[1]{\textcolor{tolpurple}{\oldref{#1}}}

\newcommand{\red}[1]{{\color{red}#1}}
\newcommand{\blue}[1]{{\color{blue}#1}}
\newcommand{\green}[1]{{\color{green}#1}}
\newcommand{\ps}{\red}

\preprint{APS/123-QED}

\title{Impurity-dependent quantum geometry in
2D flat band superconductors}%

\author{Simão S. Cardoso}%
\email{simao.simao-silva-cardoso@universite-paris-saclay.fr}
\author{A. Mesaros}%
\email{andrej.meszaros@universite-paris-saclay.fr}
\author{P. Simon}%
\email{pascal.simon@universite-paris-saclay.fr}

\affiliation{Laboratoire de Physique des Solides (CNRS UMR 8502),\\ Bâtiment 510, Université Paris-Sud/Université Paris-Saclay, 91405 Orsay, France}%

\date{\today}

\begin{abstract}
In flat band superconductors, quantum geometry, rather than kinetic energy, governs pairing and transport. However, how this geometry manifests in the response to a local, individually addressable impurity remains largely unexplored. Here, we study the Yu-Shiba-Rusinov (YSR) bound state induced by a magnetic adatom that hybridizes with every orbital of a two-dimensional flat band superconductor, and characterize its spatial profile using two quantities: the quadratic spread \(\mathcal{Q}_S\) and the localization length \(\xi\). We show that \(\mathcal{Q}_S\) is not set by the quantum metric alone, as geometric corrections arising from inter-orbital interference and, more importantly, from the anisotropy of the adatom-lattice hybridization contribute on equal footing, making \(\mathcal{Q}_S\) strongly impurity dependent and tunable. On the other hand, \(\xi\) emerges from the projection onto the flat band alone and is a universal property set by the overlap of compact localized states. All impurity dependence is instead isolated in the prefactor of the wavefunction whose magnitude correlates directly with \(\mathcal{Q}_S\). We, therefore, report a clean separation between universal and impurity-tunable physics from the induced YSR bound state in these systems, and confirm our results numerically via exact diagonalization.
\end{abstract}

\maketitle
\paragraph*{Introduction -} Flat band (FB) systems have attracted considerable attention in both theoretical and experimental research. Many of the interesting findings are intrinsically connected to the quantum geometry encoded in the quantum geometric tensor (QGT) of Bloch states that governs fundamental electronic and topological properties \cite{Resta2011}. Its real part is the quantum metric tensor of the Bloch states, and the imaginary counterpart is the Berry curvature \cite{berry_1984}. Recently, increased attention has been devoted to the quantum metric, which measures the distance between two adjacent Bloch states \cite{provost_1980}, and has emerged as a key quantity in various physical phenomena, especially in superfluidity and superconductivity in FBs \cite{chen_ginzburg-landau_2024,peotta_superfluidity_2015,torma_2018,julku_2016,topp_light-matter_2021,herzog-arbeitman_superfluid_2022,huhtinen_2022,thumin_2023}.
\par With the discovery of a superconducting order on twisted bilayer graphene, a material with a flat Moiré band at the magic angle \cite{cao_unconventional_2018,tian_2023,lu_2019}, FBs emerge as a promising platform for enhanced critical temperatures \cite{MIYAHARA20071145}. Alongside superconductivity comes the study of impurities, as they can change the superconducting properties and lead to the creation of sub-gap bound states \cite{balatsky_2006}. Magnetic impurities, inducing Yu-Shiba-Rusinov (YSR) bound states, are a well-established, experimentally accessible probe of local pairing and normal-state structure in conventional superconductors \cite{yu_1965,shiba_1968,rusinov_1969,bauriedl_1981,yazdani_1997,Menard2015,Thupakula_2022} (see \cite{Heinrich_2018} for a review). As the main motivation behind this work, some questions are raised: what happens when the host is a FB superconductor instead of a dispersive one? How does quantum geometry show up in a local, individually addressable impurity problem? How does the spatial profile of the bound state depend on the impurity?
\par Although inhomogeneities in FB systems have been extensively studied recently \cite{ZHU2019,basak_2022,Marques2024,li2025,ktlaw_2025,guo_majorana_2025,ma_2025,Chau_2026,liu_2026,zhao_2026}, only a small number of works have addressed the connection between the spatial extent of bound states and the quantum geometry. More specifically, it has been shown that a length scale defined from the quantum metric governs, or at least sets a lower bound to the spatial decay, of the proximity effect between a superconductor and a FB material \cite{ktlaw_2025}, of Majorana bound states \cite{guo_majorana_2025}, and FB topological boundary modes \cite{ma_2025}. However, due to the embedding dependence, a more universal localization length was proposed, determined by compact localized states (CLS) overlaps \cite{lee_embedding_2025,kim2026}. The algebraic properties of CLS determine the long-distance behavior of the FB projector in real space, which sets an upper bound for the decay of a sub-gap bound state. 
\par Furthermore, the study of more complex and realistic impurities, such as adatoms, and their associated quantum geometric signatures in the spread of the bound state remains largely unexplored. Intuitively, one might expect the real-space spread of the bound state to be set by the quantum metric of the FB, reminiscent of the Wannier-function localization in \cite{marzari_maximally_1997}, where the minimal spread is controlled by the Brillouin Zone (BZ) averaged quantum metric, a theoretical prediction recently made for Majorana zero modes in FBs \cite{ktlaw_2025}. However, with the introduction of an adatom hybridizing with the orbitals in the unit cell, the existing models are insufficient because they neglect the hybridization anisotropy between the adatom and the orbitals of the unit cell, which we show introduces geometric corrections beyond the quantum metric alone. 
\par In this work, we analyze the spatial properties of magnetic adatom-induced bound states in two-dimensional FB superconductors, both analytically and numerically. Two quantities are considered: the variance of position (quadratic spread \(\mathcal{Q}_S\)), which measures the spatial extent of the bound state, and the localization length \(\xi\). For the former, we show that the quantum metric alone does not determine the spread. Hybridization anisotropy between the adatom and the orbitals in the unit cell of the host lattice contributes corrections of the same order, entering through both the second and first moments of position. Thus, \(\mathcal{Q}_S\) is strongly dependent on the impurity. For the latter, the length scale \(\xi\) arises from the projection onto the FB, being a global quantity independent of impurity properties. The developed calculations are applied to the Lieb lattice, a bipartite two-dimensional lattice with a FB that can be isolated from the dispersive ones via dimerization. Numerically diagonalizing the real-space tight-binding Hamiltonian demonstrates excellent agreement with the analytical predictions. The detailed derivation of the analytical results and further numerical simulations are provided in the supplementary material (SM).\\

\paragraph*{Flat band superconductor -} We start by considering that our system is described by the Hubbard model
\begin{align}
    \hat{H}_S = \sum_{\mathbf{R}_i\alpha,\mathbf{R}_j\beta,\sigma}&t_{ij}^{\alpha\beta}\hat{c}_{\mathbf{R}_i,\alpha,\sigma}^\dagger \hat{c}_{\mathbf{R}_j,\beta,\sigma} \nonumber\\
    &-|U|\sum_{\mathbf{R}_i,\alpha}\hat{n}_{\mathbf{R}_i,\alpha,\uparrow}\hat{n}_{\mathbf{R}_i,\alpha,\downarrow},
\end{align}
where \(\mathbf{R}_i = n_i a_1 \hat{\mathbf{e}}_x + m_i a_2\hat{\mathbf{e}}_y\) (\(n_i,m_i \in \mathbb{Z}\)) labels the position of unit cell 
\(i\) on a general 2D Bravais lattice with lattice vectors \((a_1\hat{\mathbf{e}}_x,a_2\hat{\mathbf{e}}_y)\), \(\alpha,\beta\) are the orbitals within the unit cell, and \(\sigma = \uparrow, \downarrow\) represents the spin. The first term describes the neighbor hopping via \(t_{ij}^{\alpha\beta}\), where \(\hat{c}_{\mathbf{R}_i,\alpha,\sigma}(\hat{c}_{\mathbf{R}_i,\alpha,\sigma}^\dagger)\) are the annihilation (creation) lattice operators. The chemical potential \(\mu\) is absorbed into the diagonal hopping terms \(t_{ii}^{\alpha\alpha}\) and fixed at the flat band energy, \(\mu=\epsilon_0=0\), corresponding to half-filling. The last term is the electron-electron interaction, which is a contact interaction for opposite-spin electrons located on the same orbital with \(|U|>0\), and \(\hat{n}_{\mathbf{R}_i,\alpha,\sigma}\) is the number operator. The typical BCS approximation is obtained by applying the mean-field decoupling scheme to the Hubbard interaction term, where the pairing \(\Delta_\alpha\) is \(s\)-wave and calculated self-consistently as \(\Delta_\alpha = -|U| \langle \hat{c}_{\mathbf{R}_i,\alpha,\downarrow}\hat{c}_{\mathbf{R}_i,\alpha,\uparrow}\rangle\). Moreover, we assume orbital-independent pairing \(\Delta_\alpha = \Delta\).
\par Moving to \(\mathbf{k}\)-space allows to write the Hamiltonian
\begin{align}
\label{eq:superconductor_mag}
    \hat{H}_S
= \sum_{\mathbf{k},\sigma,\alpha,\beta}&h_{\mathbf{k}}^{\alpha \beta}
\hat{c}^\dagger_{\mathbf{k},\alpha,\sigma}
\hat{c}_{\mathbf{k},\beta,\sigma} \nonumber\\
&+ \Delta \sum_{\mathbf{k},\alpha}
\left(
\hat{c}^\dagger_{\mathbf{k},\alpha,\uparrow}\hat{c}^\dagger_{-\mathbf{k},\alpha,\downarrow}+
\text{h.c.}\right).
\end{align}
The lattice is represented by the Bloch Hamiltonian \(h_\mathbf{k}\) with dimension \(\mathcal{N}_\alpha\times \mathcal{N}_\alpha\), where \(\mathcal{N}_\alpha\) is the number of orbitals per unit cell. After diagonalization of the Bloch Hamiltonian, we obtain the energies of the bands \(\epsilon_n(\mathbf{k})\) with \(n\) the band index, and their respective Bloch states \(\ket{u_n(\mathbf{k})}\). Our focus is on the FB Bloch state \(\ket{u_0(\mathbf{k})}\), which obeys the relation \(\sum_{\alpha,\beta}h_\mathbf{k}^{\alpha \beta}u_0^\beta(\mathbf{k}) = 0\).
\par
\begin{figure}
\centerline{\includegraphics[width=0.48\textwidth]{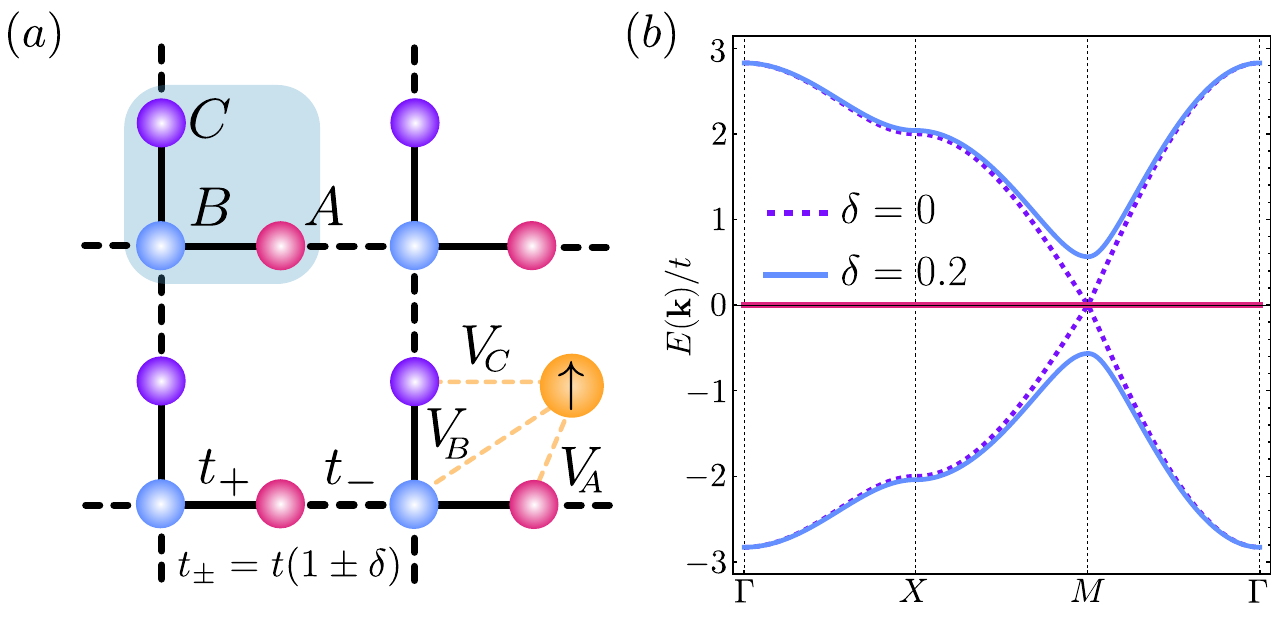}}
\caption{(a) Lieb lattice with the three orbitals labeled as \(A,B,C\). The intra-cell hopping \(t_+\) is represented by the thick black lines, while the inter-cell hopping \(t_-\) is shown by the black dashed lines. The amber orbital is the magnetic adatom, and hybridizes to every orbital with the hybridization vector \(\ket{\mathcal{V}}=(V_A,V_B,V_C)^T\). (b) Spectrum of the normal-state Lieb lattice with the FB represented by the red line. When \(\delta=0\), the dispersive bands touch the FB at the high symmetry point \(M = (\pi/a,\pi/a)\), in purple dashed lines. When \(\delta \neq 0\), a gap opens between the bands, as shown by the blue lines.}
\label{fig:1}
\end{figure}

 For this work, we turn to the Lieb lattice, a decorated square lattice with three orbitals per unit cell \(A,B,C\), partitioned into sublattices \(\mathcal{A}=\{A,C\}\), and \(\mathcal{B}=\{B\}\), with \(B\) acting as the hub, as displayed schematically in Fig.~\ref{fig:1}(a). The sublattice imbalance of this bipartite lattice pins a zero-energy FB, which can be isolated from the remaining dispersive bands by a hopping dimerization. The FB eigenstates vanish on the hub \(B\) through destructive interference, carrying weight solely on the sublattice \(\mathcal{A}\). While a self-consistent solution yields a nonzero hub pairing \(\Delta_B\) that differs from those on \(\mathcal{A}\), it is irrelevant to the FB physics precisely because these eigenstates have no support on \(B\). The \(C_4\) rotation about the hub exchanges \(A\) and \(C\) enforcing \(\Delta_A=\Delta_C\) \cite{julku_2016}, so the FB pairing is governed by the single amplitude \(\Delta \equiv \Delta_A=\Delta_C\), which we treat as isotropic in what follows.
\par The normal-state Bloch Hamiltonian that defines the Lieb lattice is expressed below in unit-cell gauge as
\begin{align}
\label{eq:sutb_bloch_hamiltonian}
    h_\mathbf{k} &= \begin{pmatrix}
      0 & \alpha_\mathbf{k}^*  & 0 \\
      \alpha_\mathbf{k} & 0 & \beta_\mathbf{k}\\
      0 & \beta_\mathbf{k}^*  & 0
    \end{pmatrix},
\end{align}
where \(\alpha_\mathbf{k}=t_++t_- e^{-i k_x a}\), and \(\beta_\mathbf{k}=t_++t_- e^{-i k_y a}\), and \(a\) the unit cell constant. The intra-cell and inter-cell hoppings are given as \(t_\pm=t(1\pm\delta)\), respectively, leading to the dimerization of the lattice. Direct diagonalization of the Bloch Hamiltonian, Eq.~(\ref{eq:sutb_bloch_hamiltonian}), yields the spectrum observed in Fig.~\ref{fig:1}(b). A gap opens between the FB and the other dispersive bands and is controlled by the dimensionless dimerization parameter \(\delta\), with energy \(E _\text{gap} = 2\sqrt{2}\delta t\). With the introduction of superconductivity, we can express the respective Bogoliubov-de Gennes (BdG) Hamiltonian as \(H_\text{BdG}(\mathbf{k}) = \tau_z\otimes h_\mathbf{k} + \tau_x\otimes \Delta I_{\mathcal{N}_\alpha}\), with the Pauli matrices \(\tau_i\) acting in the particle-hole subspace. Due to the isotropic pairing, \(H_\text{BdG}\) is diagonal in the orbital sector of the pairing term.\\
\paragraph*{Low-energy approximation -} Due to the energy gap, \(E_\text{gap}\), which we take to be the largest energy scale in the system, the relevant physical properties remain on the FB. Therefore, we apply a low-energy approximation to the two superconducting FBs via non-square projectors \cite{guo_majorana_2025}, such that \(W_0^\dagger(\mathbf{k}) H_\text{BdG}(\mathbf{k})W_0 (\mathbf{k}) = E_0\tau_z\), and \(W_0(\mathbf{k}) =(g_+(\mathbf{k}),g_-(\mathbf{k}))\), where \(E_0 = \Delta\) is the superconducting FB energy. This step of projecting the system into the two superconducting FBs leads to the bare Green's function 
\begin{equation}
  \label{eq:bare_green}
    G_0^\text{Flat}(\mathbf{r},E) = \frac{1}{N} \sum_{\mathbf{k},n=\pm}e^{i \mathbf{k}\cdot\mathbf{r}}\frac{g_n(\mathbf{k})g_n ^\dagger (\mathbf{k})}{E-E_n},
\end{equation}
where \(E_\pm = \pm E_0=\pm  \Delta\), \(N\) is the number of \(\mathbf{k}\) points, and \(g_n(\mathbf{k})=v_n\otimes u_0(\mathbf{k})\) are the quasiparticle Bloch bands with \(v_\pm =1/\sqrt{2}(1,\pm 1)^T\) the eigenvectors of \(\Delta \tau_x\).\\

\paragraph*{Magnetic adatom -}We adopt a one-body model to describe a proximitized single magnetic impurity that hybridizes with the FB \(s\)-wave superconductor, Eq.~(\ref{eq:superconductor_mag}), \cite{machida_1972,bernevig_2018}. Comprising only a single site at the unit cell \(\mathbf{R}_i=\mathbf{0}\) and two spins, the adatom impurity is defined in real space as
\begin{equation}
    \label{eq:adatom_mag}
    \hat{H}_\text{imp} =  \sum_{\sigma,\sigma'} \hat{d}^\dagger_{\sigma}(\mathbf{J}\cdot\bm{\sigma})_{\sigma\sigma'}\hat{d}_{\sigma'}+ \sum_{\alpha,\sigma}V_\alpha\left(\hat{c}^\dagger_{0,\alpha,\sigma}\hat{d}_{\sigma} + \hat{d}^\dagger_{\sigma}\hat{c}_{0,\alpha,\sigma}\right),
\end{equation}
where \(\mathbf{J}\) is the exchange energy vector, and \(\bm{\sigma} =(\sigma_x,\sigma_y,\sigma_z)\) is the Pauli matrix vector. The operators \(\hat{d}^\dagger_{\sigma}(\hat{d}_{\sigma})\) create (annihilate) a particle at the adatom orbital. The impurity is described as a classical object: to exploit the spin-rotation symmetry of the substrate, without loss of generality, we choose the spin quantization axis of the impurity along \(z\)-axis \(\mathbf{J}=J \hat{\mathbf{e}}_z\), such that the magnetic term is proportional to the Pauli matrix \(\sigma_z\). Consequently, the Hamiltonian block-diagonalizes into independent spin sectors, allowing the problem to be solved separately within each spin subspace \cite{pientka_2013}. The second term describes tunneling via the hybridization energy \(V_\alpha\), taken to be real and momentum-independent, coupling the adatom to every orbital \(\alpha\) in the unit cell at the origin.
\par The magnetic adatom is treated as local and situated at the center of the lattice \(\mathbf{r}=\mathbf{0}\), coinciding with the unit cell 
\(\mathbf{R}_i = \mathbf{0}\). Integrating out the impurity degrees of freedom yields an effective local self-energy for the lattice electrons, encoding the hybridization with the adatom, given as
\begin{align}
    \label{eq:effective_pot_mag}
    U _\text{imp}^\sigma(E) = (E-\sigma J)^{-1}\tau_0\otimes \ket{\mathcal{V}}\bra{\mathcal{V}},
\end{align}
where \(E\) is the bound state energy, and \(\ket{\mathcal{V}}=(V_A,V_B,V_C)^T\) is the hybridization vector. The first matrix \(\tau_0\) acts as an identity in the particle-hole subspace, while the matrix \(\ket{\mathcal{V}}\bra{\mathcal{V}}\) acts in the lattice subspace. To quantify the anisotropy, we define a
hybridization strength, such that \(\bar{V} =\sqrt{\sum_\alpha V_\alpha^2/\mathcal{N}_\alpha}\). Thus, the hybridization vector becomes \(\ket{\mathcal{V}}=\bar{V}\ket{m}\), where the dimensionless vector \(\ket{m}\) can be split into orthogonal isotropic and anisotropic channels, respectively, \(\ket{m} = c \ket{s}+\ket{\eta}\), with \(\ket{s} = (1,1,1)^T\), and \(c =\braket{s|m}/\braket{s|s}\) being the isotropic weight. By expressing \(\ket{m} = \ket{\mathcal{V}}/\bar{V}\), the anisotropic channel is found as \(\ket{\eta} = \ket{\mathcal{V}}/\bar{V}-c\ket{s}\). 
\par With this definition, the new hybridization matrix is defined as \(\ket{\mathcal{V}}\bra{\mathcal{V}} = \bar{V}^2(c\ket{s}+\ket{\eta})(c\bra{s}+\bra{\eta}) \), and consequently, the local self-energy becomes
\begin{align}
    \label{eq:effective_pot_mag_2}
    U _\text{imp}^\sigma(E) = \frac{\bar{V}^2}{E-\sigma J}\tau_0\otimes \mathcal{H}_V,
\end{align}
where \(\mathcal{H}_V = c^2S +\Lambda\) with \(S= \ket{s}\bra{s}\) the isotropic matrix, and \(\Lambda = c\ket{s}\bra{\eta}+c\ket{\eta}\bra{s} + \ket{\eta}\bra{\eta}\), which isolates the anisotropic linear and quadratic contributions.\\

\begin{figure}
\centerline{\includegraphics[width=0.48\textwidth]{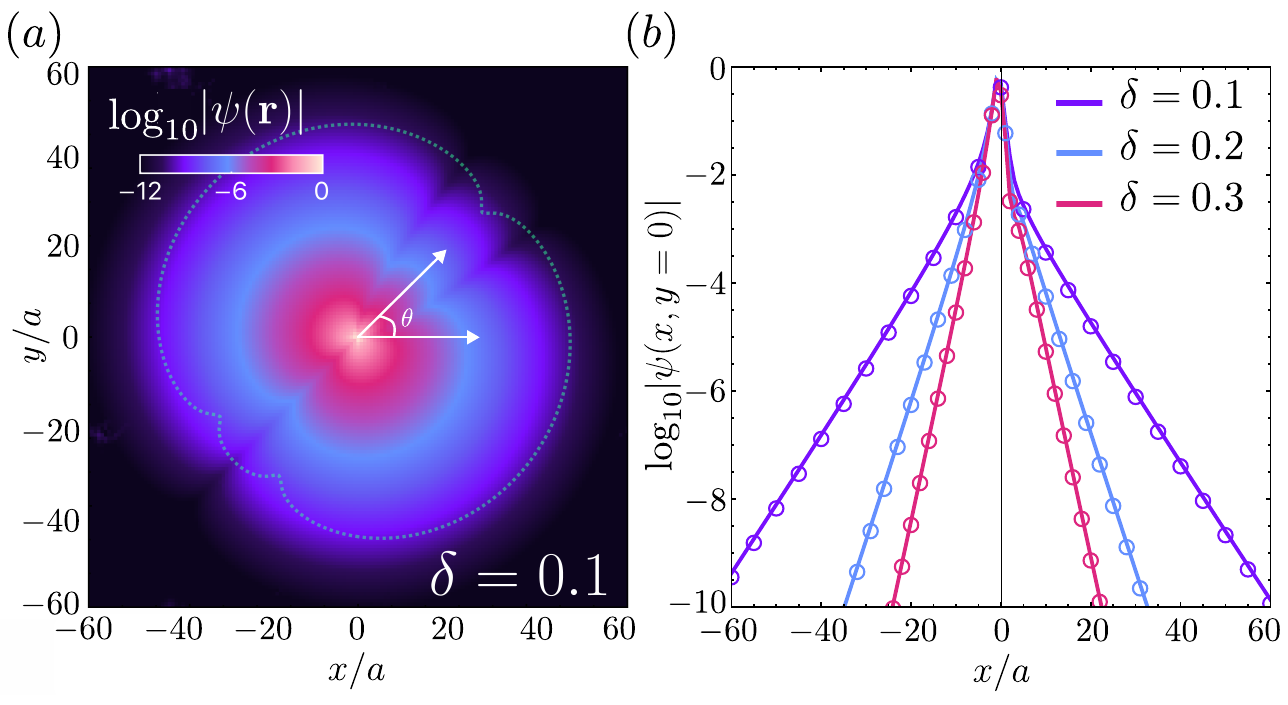}}
\caption{(a) Numerically obtained bound state density distribution \(|\psi(\mathbf{r})|\) at \(\delta=0.1\) induced by the hybridization vector \(\ket{\mathcal{V}}_\text{max}=2\Delta(1,0,1)^T\). The green dashed outline shows the respective \(|\psi(\mathbf{r})|\) at \(\ket{\mathcal{V}}_\text{min}=2\Delta(1,0,-1)^T\).  (b) Cross-section of along \(y=0\), or equivalently \(|\psi(r,\theta = 0)\) in polar coordinates, from \(\ket{\mathcal{V}}_\text{max}\) for different values of \(\delta\). The dots are obtained numerically, while the lines are computed analytically from Eq.~(\ref{eq:total_density_spin_normalized}). The parameters used are \(\Delta=1\), \(t=10^4\Delta\), \(J=60\Delta\), and \(a=1\) for a system with \(N_x=N_y=161\) unit cells (only shown \(N_x=N_y=121\) in both panels).}
\label{fig:2}
\end{figure}

\paragraph*{Bound state wavefunction -}The presence of an adatom on a FB superconductor induces a sub-gap bound state under certain conditions (discussed in SM), whose spatial profile is shown in Fig.~\ref{fig:2}(a), and its cross-section along \(y=0\) or \(|\psi(r,\theta=0)|\) in polar coordinates in Fig.~\ref{fig:2}(b). The center of the bound state is localized around the impurity at \(\mathbf{r}=\mathbf{0}\), showing three visible dips (explored via the prefactor 
\(\mathcal{A}_\alpha(\theta)\) in further sections). From the cross-sections, we find asymmetry around the center, which is quantified by Eq.~(\ref{first_moment}). Moreover, the core, within 
\(\pm 5\) unit cells of the impurity, exhibits a faster decay than the asymptotic Ornstein-Zernike (OZ) tail, Eq.~(\ref{eq:saddle_wave}), which only becomes accurate at larger \(r\). These tails have the same length scale, Eq.~(\ref{eq:xi}), on both the positive and negative sides, such that \(\xi(\theta) = \xi(\theta+\pi)\).
\par A bound state wavefunction can be expressed in Lippmann–Schwinger form, where the lattice Green’s function dresses the impurity potential, such that \(\psi_\sigma (\mathbf{r}) = G_0^\text{Flat}(\mathbf{r},E)U _\text{imp}^\sigma (E)\psi_\sigma(\mathbf{0})\) \cite{ktlaw_2025,lee_embedding_2025,lippmann_1950,pientka_2013}. With the implementation of the FB bare propagator, Eq.~(\ref{eq:bare_green}), and the effective impurity energy, Eq.~(\ref{eq:effective_pot_mag}), and combining the bound state wavefunction for each spin \(\sigma\) into the total spinor \(\psi(\mathbf{r})=(\psi_\uparrow(\mathbf{r}),\psi_\downarrow(\mathbf{r}))^T\), we obtain the total density distribution \(|\psi(\mathbf{r})|^2=|\psi_\uparrow(\mathbf{r})|^2+|\psi_\downarrow(\mathbf{r})|^2\), which becomes explicitly 
\begin{align}
    \label{eq:total_density_spin_normalized}
    |\psi(\mathbf{r})|^2 &= \frac{1}{\lambda_l}\frac{1}{N^2}\sum_{\mathbf{k},\mathbf{k}'}e^{i (\mathbf{k}-\mathbf{k}')\cdot\mathbf{r}} \mathcal{F}(\mathbf{k},\mathbf{k'}),
\end{align}
with
\begin{align}
\label{normalization}
    \mathcal{F}(\mathbf{k},\mathbf{k}') = \text{Tr}( P_{\mathbf{k}'}P_\mathbf{k}\mathcal{H}_V), 
    &&
    \lambda_l = \langle \text{Tr}( P_\mathbf{k}\mathcal{H}_V)\rangle_\text{BZ}.
\end{align}
where \(P_\mathbf{k}=\ket{u_0(\mathbf{k})}\bra{u_0(\mathbf{k})}\) is the FB projector, and we introduce the notation \(\langle A\rangle_\text{BZ} = 1/N \sum_\mathbf{k}A(\mathbf{k})\) meaning an average over the BZ. \(\lambda_l\) is a normalization constant, which can be understood as the expectation value of the hybridization matrix \(\mathcal{H}_V\) at the impurity site. Here, 
\(\text{Tr}(\cdot)\) denotes the trace over the respective subspace, while in further sections \(\text{tr}(\cdot) = \sum_{\mu}(\cdot)_{\mu\mu}\) denotes the sum over the Cartesian directions \(\mu=x,y\).
\par The total density distribution is independent of any energy scales of the impurity, depending only on the hybridization matrix, which is confirmed against numerical simulations in Fig.~\ref{fig:2}(b)\\

\paragraph*{Quadratic spread -} From Fig.~\ref{fig:2}(b), the bound state wavefunction becomes more localized as the dimerization increases. This effect can be quantified by the variance of position \(\text{Var}(\mathbf{r}) = \langle \mathbf{r}^2\rangle-\langle \mathbf{r}\rangle^2 \equiv \mathcal{Q}_S\), also known in the literature as the quadratic spread. It was first introduced in \cite{marzari_maximally_1997} as a way to localize Wannier functions maximally. It  was later used to determine the bounds for the optical spectral weight for FB superconductors \cite{randeria_2021}, and then applied to connect the BZ-averaged quantum metric either with the spread of Majorana Zero Modes in topological 1D topological FB superconductor \cite{guo_majorana_2025} or the spread of Friedel oscillations in FB materials \cite{ma_2026}. In what follows, we compute the quadratic spread by referring to the total density distribution, Eq.~(\ref{eq:total_density_spin_normalized}). We can split the hybridization matrix into three components \(\mathcal{H}_V = c^2 I + c^2 \mathcal{O} + \Lambda\), where \(\mathcal{O}=S-I\) encodes the off-diagonal elements of \(S\). Importing this result into the definition of quadratic spread, we obtain for each term 
\begin{align}
\label{second_moment}
    \langle \mathbf{r}^2\rangle &= \frac{1}{\lambda_l}\frac{1}{N}\sum_{\mathbf{k}}\left(2c^2 \text{tr}g(\mathbf{k}) + c^2\text{tr}\tilde{g}_\mathcal{O}(\mathbf{k})+\text{tr}\tilde{g}_\Lambda(\mathbf{k})\right), \\
    \label{first_moment}
    \langle \mathbf{r}\rangle &=-\frac{1}{\lambda_l}\frac{1}{N}\sum_{\mathbf{k}}\text{Im}\text{Tr}(P_\mathbf{k}\bm{\nabla}_\mathbf{k}P_\mathbf{k}(c^2\mathcal{O}+\Lambda)),
\end{align}
where we define
\begin{align}
    \label{quantum_metric}
    g_{\mu \mu}(\mathbf{k}) &= \frac{1}{2}\text{Tr}(\partial_{k_\mu}P_\mathbf{k}\partial_{k_\mu}P_\mathbf{k}), \\
    \label{isotropic_geo_correction}
    \tilde{g}_{A,\mu\mu}(\mathbf{k}) &= \text{Tr}(\partial_{k_\mu}P_\mathbf{k}\partial_{k_\mu}P_\mathbf{k}A).
\end{align}
The first term, Eq.~(\ref{quantum_metric}), is the FB quantum metric tensor, which is the real part of the QGT. In contrast, the second, Eq.~(\ref{isotropic_geo_correction}), is a geometric correction tensor for \(A=\mathcal{O},\Lambda\).
\par The quadratic spread is not captured by the quantum metric alone. The hybridization between the adatom and the host enters the formulation in two distinct ways: through the geometric corrections \(\tilde{g}_{\mathcal{O}}(\mathbf{k}),\tilde{g}_{\Lambda}(\mathbf{k})\), which supplement the quantum metric in the second moment of position \(\langle\mathbf{r}^2\rangle\), while a finite first moment \(\langle\mathbf{r}\rangle\) further subtracts from the quadratic spread. These inter-orbital interference effects in \(\mathcal{O}\) and anisotropic corrections in \(\Lambda\) contribute to the asymmetry of the bound state wavefunction, as observed in Figs.~\ref{fig:2}(a,b).\\

\paragraph*{Impurity dependence of the quadratic spread -} To better understand how anisotropy in the hybridization between the adatom and the lattice, given by \(\ket{\mathcal{V}}=(V_A,V_B,V_C)^T\), affects the quadratic spread of the bound state, we convert the problem into spherical coordinates, such that \(\ket{\mathcal{V}} = \sqrt{3}\bar{V}(\sin\theta\cos\phi,\sin\theta\sin\phi,\cos\theta)\), with \(\theta \in [0,\pi]\), and \(\phi\in [0,2\pi]\) the spherical angles. Moreover, the anisotropic channel becomes \(\ket{\eta} = \ket{\mathcal{V}}/\bar{V}-c(\theta,\phi)\ket{s}\), with \(c(\theta,\phi) = (\sin\theta\cos\phi+\sin\theta\sin\phi+\cos\theta)/\sqrt{3}\).
\begin{figure}
\centerline{\includegraphics[width=0.48\textwidth]{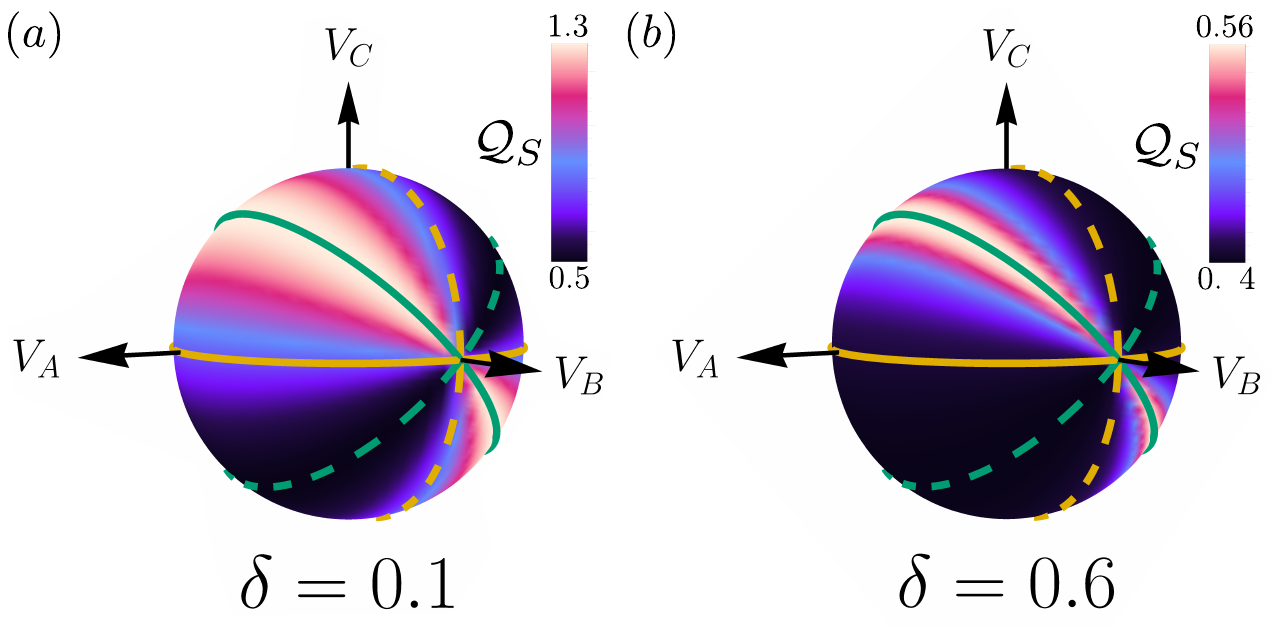}}
\caption{Quadratic spread mapped onto the surface of a sphere with radius \(\sqrt{3}\bar{V}\), Eqs.~(\ref{second_moment},\ref{first_moment}), for every possible anisotropy \(\ket{\mathcal{V}}=(V_A,V_B,V_C)^T\) at (a) \(\delta=0.1\), and (b) \(\delta=0.6\). The yellow lines, dashed and full, represent the anisotropies \(\ket{\mathcal{V}}=(0,V_B,V_C)^T\), and \(\ket{\mathcal{V}}=(V_A,V_B,0)^T\), while the full and dashed green lines exhibit the anisotropies \(\ket{\mathcal{V}} = (V_A,V_B,V_A)^T\), and \(\ket{\mathcal{V}} = (V_A,V_B,-V_A)^T\), respectively. Because the FB weight is only carried in orbitals \(A,C\), the quadratic spread remains unchanged for whatever value of \(V_B\). Thus, we observe the same behavior for \(-V_B\). It was assumed \(a=1\).}
\label{fig:3}
\end{figure} 
\par Updating the hybridization matrix \(\mathcal{H}_V\rightarrow \mathcal{H}_V(\theta,\phi)\) allows us to map the quadratic spread over every possible anisotropy, parametrized as points on the surface of a sphere of radius \(\sqrt{3}\bar{V}\), which is shown in Figs.~\ref{fig:3}(a,b). The anisotropies \(\ket{\mathcal{V}}=(0,V_B,V_C)^T\), and \(\ket{\mathcal{V}}=(V_A,V_B,0)^T\), shown by the yellow dashed and full lines, respectively, yield the same quadratic spread as a function of coupling strength, since the \(A \leftrightarrow C\) exchange (the \(C_4\) rotation about the hub) maps one orbital onto the other. On the other hand, the anisotropies \(\ket{\mathcal{V}} = (V_A,V_B,V_A)^T\), and \(\ket{\mathcal{V}} = (V_A,V_B,-V_A)^T\), represented by the green full and dashed lines, respectively, provide the maximal and minimal spread for the Lieb lattice, a feature we find to hold independent of dimerization, as seen in Figs.~\ref{fig:3}(a,b). We therefore define \(\ket{\mathcal{V}} = (V_A,V_B,V_A)^T \equiv \ket{\mathcal{V}}_\text{max}\), and \(\ket{\mathcal{V}} = (V_A,V_B,-V_A)^T \equiv \ket{\mathcal{V}}_\text{min}\) for later sections in this work.
\begin{figure}
\centerline{\includegraphics[width=0.48\textwidth]{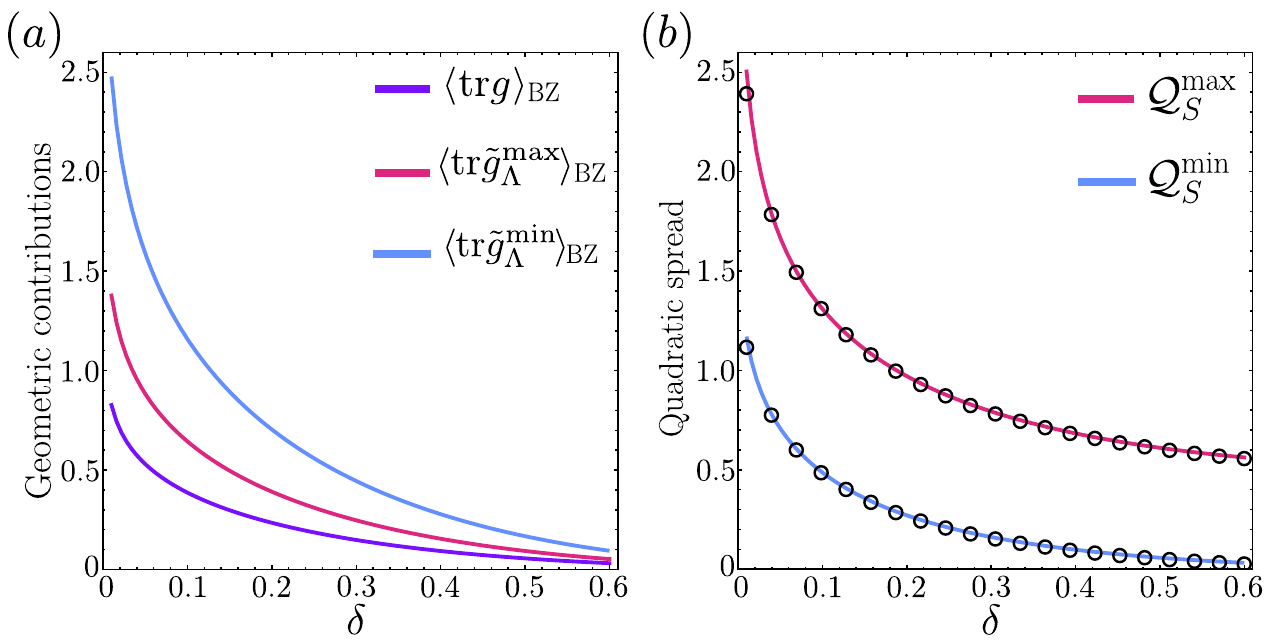}}
\caption{(a) Geometric contributions present in \(\langle \mathbf{r}^2\rangle\) averaged over the BZ as a function of the dimerization parameter \(\delta\). The purple line shows the quantum metric tensor, while the anisotropic corrections for both \(\mathcal{Q}_S^\text{max}\), and \(\mathcal{Q}_S^\text{min}\) are shown in red and blue, respectively. (b)
Quadratic spread as a function of the dimerization parameter \(\delta\) for both anisotropic regimes. The analytical results are shown in red and blue lines, while the black circles were obtained numerically from the diagonalization of the
real-space tight-binding Hamiltonian (\(N_x = N_y = 91\) unit cells) with the magnetic adatom. For the maximal quadratic spread regime it was used \(\ket{\mathcal{V}}_\text{max}=2\Delta(1,0,1)^T\), and for the minimal regime \(\ket{\mathcal{V}}_\text{min}=2\Delta(1,0,-1)^T\). The parameters used are \(\Delta=1\), \(t=10^4\Delta\), \(J=60\Delta\), and \(a=1\).}
\label{fig:4}
\end{figure} 
\par To get a clearer picture of the quadratic spread formulation analytically, we turn our attention to the second moment of position, Eq.~(\ref{second_moment}), which is the main contributor to \(\mathcal{Q}_S\). Besides the quantum metric, the first correction term \(\tilde{g}_\mathcal{O}(\mathbf{k})\) vanishes for the Lieb lattice because \(u_0(\mathbf{k})\) has no support on the hub orbital \(B\), killing the cross terms in \(\mathcal{O}\). On the other hand, the anisotropic correction \(\tilde{g}_\Lambda\), by construction, scales linearly with the anisotropic hybridization matrix \(\Lambda\). In Fig.~\ref{fig:4}(a), both anisotropic corrections \(\langle \text{tr}\tilde{g}_{\Lambda}^\text{max}\rangle_\text{BZ},\langle \text{tr}\tilde{g}_{\Lambda}^\text{min}\rangle_\text{BZ} \gtrsim 2\langle \text{tr}g\rangle_\text{BZ}\).
\par In Fig.~\ref{fig:4}(b), we provide the complete variation of the quadratic spread over the dimerization for both anisotropic regimes \(\ket{\mathcal{V}}_\text{max}\rightarrow \mathcal{Q}_S^\text{max}\), and \(\ket{\mathcal{V}}_\text{min}\rightarrow \mathcal{Q}_S^\text{min}\). A numerical simulation is achieved by exact diagonalization of a real-space tight-binding Hamiltonian of the Lieb lattice with \(N_x\times N_y\) unit cells, periodic boundary conditions, and with the magnetic adatom localized at the center of the lattice with coordinates \((N_x/2,N_y/2)\). From the induced sub-gap bound state obtained, we extract the corresponding eigenstate and compute its quadratic spread (more details in SM). The comparison with the numerical simulation, given by the black circles in Fig.~\ref{fig:4}(b), reveals a perfect match with our analytical results in both anisotropic regimes.\\

\paragraph*{Impurity independence of the localization length  -} The induced bound state wavefunction, obtained numerically, is represented spatially in Fig.~\ref{fig:2}(a) with the magnetic adatom centered in \(\mathbf{r} = \mathbf{0}\). The bound state shown is produced by anisotropy that provides the maximal spread \(\ket{\mathcal{V}}_\text{max} = 2\Delta(1,0,1)^T\), while the green dashed outline represents the case of \(\ket{\mathcal{V}}_\text{min} = 2\Delta(1,0,-1)^T\). Because the FB is gapped from the dispersive bands, it belongs to the class of \emph{linearly independent FBs}, characterized by exponentially decaying projectors with algebraic prefactors due to the finite overlap of neighboring CLSs \cite{kim2026}. We recall that the bound state wavefunction is obtained as \(\psi_\sigma (\mathbf{r}) = G_0^\text{Flat}(\mathbf{r},E)U _\text{imp}^\sigma (E)\psi_\sigma(\mathbf{0})\). The asymptotic spatial decay at large distances is governed entirely by the spatial decay of the projector \(P_\mathbf{k} = \ket{u_0(\mathbf{k})}\bra{u_0(\mathbf{k})}\) in the Green's function \(G_0^\text{Flat}(\mathbf{r},E) \propto 1/N\sum_\mathbf{k}e^{i\mathbf{k}\cdot\mathbf{r}}P_\mathbf{k}\), Eq.~(\ref{eq:bare_green}), \cite{lee_embedding_2025}. Thus, the bound state wavefunction can be expressed in the Ornstein-Zernike (OZ) form for an orbital \(\alpha\) with a saddle point analysis (see SM for a detailed derivation) as
\begin{align}
    \label{eq:saddle_wave}
    |\psi_\alpha(\mathbf{r})| \propto \mathcal{A}_\alpha(\theta) r^{-1/2}  e^{-r/\xi(\theta)},
\end{align}
where \(\mathcal{A}_\alpha(\theta)\) is an angle-dependent prefactor (full expression shown in SM), and \(\xi(\theta)\) is the localization length obtained as
\begin{align}
    \label{eq:xi}
    \xi(\theta) =a\bigg[\text{arcsinh}&\left(\gamma(\theta)\cos{\theta}\right)\cos{\theta} \nonumber\\
&+\text{arcsinh}\left(\gamma(\theta)\sin{\theta}\right)\sin{\theta}\bigg]^{-1},
\end{align}
with
 \begin{align}
     \label{eq:lambda}
     \gamma^2(\theta) &= \frac{2\rho^2-\rho\sqrt{2} \sqrt{\rho^2+4-(\rho^2-4)\cos{(4\theta)}}}{2 \cos^2{(2\theta)}},
 \end{align}
where \(\rho = 2(1+\delta^2)/(1-\delta^2)\). The polar angle \(\theta\) is represented in Fig.~\ref{fig:2}(a). Notably, Eqs.~(\ref{eq:xi},\ref{eq:lambda}) contain no dependence on the impurity parameters \(\bar{V}\), \(J\), \(\mathcal{H}_V\), meaning \(\xi(\theta)\) is fixed entirely by the lattice dimerization \(\delta\), and direction \(\theta\). When the gap closes, \(\delta \rightarrow 0\), then \(\gamma(\theta) \rightarrow 0\), yielding \(\xi(\theta) \rightarrow \infty\). As the FB touches the dispersive bands, the low-energy projection breaks down, and the bound state should delocalize completely.
\begin{figure}
\centerline{\includegraphics[width=0.48\textwidth]{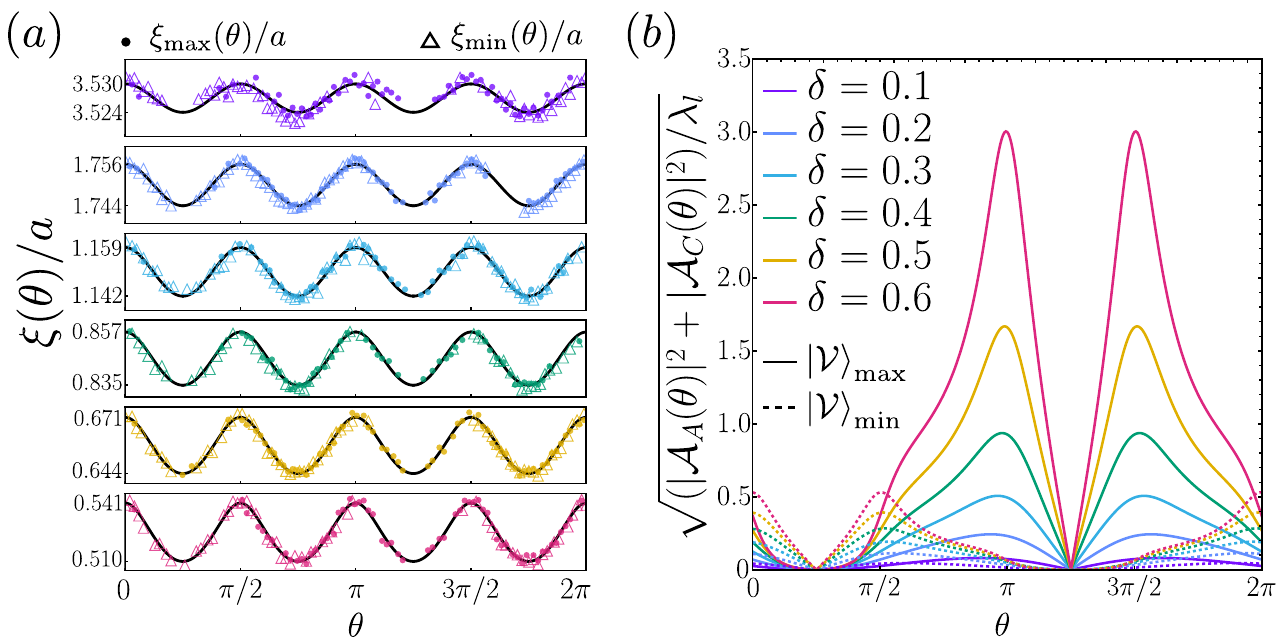}}
\caption{(a) Numerical fitting of the localization length \(\xi(\theta)\), represented for both anisotropic regimes in dots as \(\xi_\text{max}(\theta)\), and in triangles for \(\xi_\text{min}(\theta)\), while the black line is obtained analytically from Eq.~(\ref{eq:xi}). (b) Total prefactor \(\sqrt{(|\mathcal{A}_A(\theta)|^2+|\mathcal{A}_C(\theta)|^2)/\lambda_l}\) as a function of the polar angle \(\theta\) for both anisotropic regimes \(\ket{\mathcal{V}}_\text{max}=2\Delta(1,0,1)^T\), and \(\ket{\mathcal{V}}_\text{min}=2\Delta(1,0,-1)^T\), in full and dashed lines, respectively. The numerical results were computed from the diagonalization of the real-space tight-binding Hamiltonian (\(N_x=N_y=91\) unit cells) with the magnetic adatom. The parameters used are \(\Delta=1\), \(t=10^4\Delta\), \(J=60\Delta\), and \(a=1\).}
\label{fig:5}
\end{figure} 
\par Eq.~(\ref{eq:xi}) is confirmed numerically by cutting cross-sections at fixed values of \(\theta \in [0,2\pi]\) of the bound state wavefunction, as seen in Fig.~\ref{fig:2}(b), and fitting the tails using the OZ as an ansatz, Eq.~(\ref{eq:saddle_wave}), excluding around the angular dips discussed below (more detail in SM). The extracted numerical localization length as a function of \(\theta\) is in excellent agreement with Eq.~(\ref{eq:xi}), observed in Fig.~\ref{fig:5}(a), across both anisotropic opposite regimes \(\ket{\mathcal{V}}_\text{max}\rightarrow\xi_\text{max}\), and \(\ket{\mathcal{V}}_\text{min}\rightarrow\xi_\text{min}\), and is independent of the impurity parameters. Therefore, \(\xi(\theta)\) is a universal property of the host lattice, unaffected by how the adatom couples to the orbitals in the unit cell.
\par All impurity dependence is instead found in the prefactor \(\mathcal{A}_\alpha(\theta)\), which vanishes for certain values of 
\(\theta\) dictated by lattice symmetry. This, in turn, produces the angular dips seen in the bound state wavefunction, Fig.~\ref{fig:2}(a), that match the angles \(\theta\) where the prefactor vanishes in Fig.~\ref{fig:5}(b) for both \(\ket{\mathcal{V}}_\text{max}\) and  \(\ket{\mathcal{V}}_\text{min}\). The total probability weight at large \(\mathbf{r}\) is set by \(|\mathcal{A}_A(\theta)|^2+|\mathcal{A}_C(\theta)|^2\), normalized by \(\lambda_l\), while \(\xi(\theta)\) remains fixed. A larger combined prefactor, therefore, increases \(\langle \mathbf{r}^2\rangle\) and hence \(\mathcal{Q}_S\), which is the mechanism underlying the correspondence between prefactor magnitude, Fig.~\ref{fig:5}(b), and quadratic spread, Figs.~\ref{fig:3}(a,b), observed in both anisotropic regimes.\\

\paragraph*{Conclusion -} Returning to the questions posed in the introduction, we analyzed the spatial profile of a magnetic adatom-induced YSR bound state in the superconducting Lieb lattice. 
To characterize the spatial spread of the bound-state wavefunction and its sensitivity to the underlying quantum geometry, we introduced the quadratic spread \(\mathcal{Q}_S\). The quantum metric, though essential, is one contribution among other geometric corrections: \(\tilde{g}_\mathcal{O}(\mathbf{k})\) probing inter-orbital interference (which vanishes for the Lieb lattice), and \(\tilde{g}_\Lambda(\mathbf{k})\), encapsulating the anisotropic hybridization between the adatom and the host lattice. The latter turns the quadratic spread into a strongly impurity-dependent quantity. Depending on the geometry of the host lattice, the hybridization anisotropy can serve as a tuning mechanism for controlling state localization.
\par On the other hand, to evaluate the real-space decay of the bound state wavefunction, we computed from saddle point analysis the localization length \(\xi(\theta)\). This quantity is direction-dependent but global in the sense that it depends only on properties of the lattice and remains unaffected by different hybridization anisotropies. Representing the bound state wavefunction in OZ form, all the impurity dependence is introduced in the prefactor, whose magnitude directly sets the wavefunction's tail weight and hence correlates with the quadratic spread, linking the impurity-dependent and impurity-independent quantities of the problem.
\par Further analysis is provided in the SM. For the Lieb lattice, we examine how the bound state energy relates to the FB Bloch states and how it evolves across the superconducting gap, where it undergoes a quantum phase transition (QPT). Additionally, we provide a similar analysis to the one-dimensional Lieb chain, where the symmetry-enforcing \(A \leftrightarrow C\) is broken, and how that translates to a different impurity dependence behavior of the quadratic spread. 
\par We acknowledge M. Thumin for his insights on flat band systems, and F. Piéchon and A. Syrota for useful discussions throughout the development of this work. This work was supported by the International Research Laboratory 'Frontiers of Quantum Science' (IRL-LFQ), a joint collaboration between the Centre National de la Recherche Scientifique (CNRS) and the Université de Sherbrooke.

\bibliography{letter_bib}

\end{document}